\documentclass[sigplan,10pt,nonacm,screen]{acmart}
\usepackage{my_style}
\begin{document}

\title{GTaP: A GPU-Resident Fork-Join Task-Parallel System with a Pragma-Based Interface}

\author{Yuki Maeda}
\email{maeda@eidos.ic.i.u-tokyo.ac.jp}
\affiliation{%
  \institution{The University of Tokyo}
  \department{Department of Information and Communication}
  \city{Tokyo}
  \country{Japan}
}

\author{Kenjiro Taura}
\email{tau@eidos.ic.i.u-tokyo.ac.jp}
\affiliation{%
  \institution{The University of Tokyo}
  \department{Department of Information and Communication}
  \city{Tokyo}
  \country{Japan}
}

\begin{abstract}
Graphics Processing Units (GPUs) excel at regular data-parallel workloads.
In contrast, many irregular workloads are naturally expressed using fork-join task parallelism, which is well supported on CPUs but remains difficult to execute efficiently on GPUs.
We present \emph{GTaP}, a task-parallel programming system that executes fine-grained fork-join entirely within a persistent GPU kernel.
Programmers annotate fork and join points in task functions with OpenMP-like pragmas, and a Clang extension generates the suspension and resumption logic required at joins.
GTaP exposes a \emph{block mode}, which maps each task to one thread block for cooperative parallel execution within the task, and a \emph{thread mode}, which maps each task to one GPU thread for finer-grained parallelism.
Scheduling is fully GPU-resident via work stealing; \emph{Divergence-Aware Queueing (DAQ)} optionally partitions task queues using user-defined criteria to reduce warp divergence from heterogeneous control flow.

Across microbenchmarks, GTaP outperforms a prior GPU implementation of fine-grained fork-join by up to two orders of magnitude, and matches or exceeds OpenMP and OpenCilk on 72 CPU cores for compute-intensive workloads.
On real-world workloads, GTaP's pragma-annotated recursive implementations run up to $1.6\times$ faster than a hand-written GPU traversal for FMM dual-tree traversal, and outperform a state-of-the-art specialized kernel for $k$-clique counting by up to $5.2\times$ when search-tree skew causes load imbalance under static work assignment.

\end{abstract}

\ccsdesc[500]{Software and its engineering~Runtime environments}
\ccsdesc[300]{Software and its engineering~Parallel programming languages}

\keywords{GPU, fork-join, task parallelism, irregular applications, work stealing, runtime system, compilers}

\maketitle

\section{Introduction} \label{chap:introduction}

Graphics Processing Units (GPUs) are widely used as accelerators in domains such as scientific computing and machine learning, where they exploit massive hardware parallelism for regular data-parallel workloads.
Such workloads are commonly expressed using GPU programming models ranging from low-level APIs such as CUDA and HIP~\cite{cuda, hip} to directive-based approaches such as OpenMP target and OpenACC~\cite{openmp, openacc}.

In contrast, many important applications exhibit irregular parallelism (e.g., search, recursive decomposition, and computations with dynamic dependencies) that is difficult to express efficiently with simple data-parallel structures.
Task parallelism is a natural fit for such workloads.
Fork-join is a common task-parallel control structure in which a parent spawns child tasks (\emph{fork}) and later resumes after waiting for them to complete (\emph{join}).
While CPU runtime systems are mature~\cite{omptask, cilk, tbb, mth}, GPU execution could also benefit if tasks are mapped efficiently to GPU resources and scheduled with low overhead.

However, achieving fork-join efficiently on GPUs is challenging.
Launching a kernel for each fine-grained task is impractical due to launch and synchronization overheads, so prior work often relies on \emph{persistent kernels} that repeatedly fetch tasks on the device~\cite{softshell, whippletree, atos}.
Moreover, GPUs lack direct support for suspending a task and later resuming it at the same point, which join requires.
Finally, when multiple independent tasks are naively mapped to different threads within a warp, warp divergence reduces effective throughput.

Existing GPU task-parallel runtime systems~\cite{softshell, whippletree, atos, chatterjee2013, kiuchi2025} often have design constraints in programmability and efficiency.
In particular, to the best of our knowledge, no prior system simultaneously provides (i) GPU-resident dynamic load balancing, (ii) fork-join with in-place resumption of the parent context, and (iii) support for both thread and block execution modes.

To address this gap, we design and implement \emph{GTaP} (\emph{G}PU-resident \emph{Ta}sk \emph{P}arallelism), a task-parallel programming system that supports fork-join entirely on the GPU.
Unlike systems that schedule tasks at kernel granularity, GTaP runs many fine-grained tasks inside a single persistent kernel.
GTaP exposes two complementary task-execution modes: the \emph{block mode} assigns one task to an entire thread block, enabling cooperative parallel execution within a task, whereas the \emph{thread mode} assigns one task to each GPU thread, allowing tasks to be written in a conventional sequential style while exposing finer-grained task parallelism.
To mitigate warp divergence in thread-mode execution, GTaP provides \emph{Divergence-Aware Queueing (DAQ)}, which exposes multiple task queues and allows programmers to route tasks expected to follow the same control-flow path to the same queue.
GTaP targets NVIDIA GPUs and is implemented in CUDA C++.

Our contributions are as follows:
\begin{itemize}
  \item
    We design and implement GTaP, to our knowledge, the first GPU task-parallel programming system to combine fork-join with in-place parent resumption, GPU-resident distributed work stealing, and both block and thread execution modes.
    An OpenMP-inspired pragma-based API and Clang extension provide these capabilities without requiring programmers to manually restructure control flow.
  \item
    We develop GPU-side scheduling mechanisms for fine-grained tasks, including warp-cooperative batched queue operations that amortize work-stealing overhead across a warp, and DAQ for divergence-aware scheduling of thread-mode tasks.
  \item
    We evaluate GTaP on an NVIDIA GH200 using microbenchmarks, scheduler ablations, and two real-world irregular workloads.
    GTaP outperforms a prior fine-grained GPU fork-join runtime by up to two orders of magnitude, matches or outperforms OpenMP and OpenCilk on 72 CPU cores for compute-intensive workloads, and outperforms hand-written or specialized GPU baselines by up to $1.6\times$ on FMM dual-tree traversal and $5.2\times$ on deep, skewed $k$-clique searches.
    Our ablations isolate the contribution of each scheduling mechanism.
\end{itemize}

We release GTaP as open-source software at \url{https://github.com/yukim0359/GTaP}.

\section{Background} \label{chap:background}

\subsection{Task Parallelism and Fork-Join} \label{sec:task-parallelism-and-fork-join}

Task parallelism expresses irregular parallelism---often hard to capture with regular data-parallel constructs---as dynamically created units of work (\emph{tasks}) executed in parallel.
However, spawning only independent tasks, as in a simple \emph{bag-of-tasks} model, is often insufficient: many irregular applications require expressing dependencies among tasks.
Fork-join is a common control structure for this purpose, where a parent task spawns child tasks (\emph{fork}) and later waits at a join point until they complete (\emph{join}).

On CPUs, many fork-join task-parallel systems have been developed: language-integrated systems such as Cilk~\cite{cilk}, runtime systems such as Intel TBB~\cite{tbb} and MassiveThreads~\cite{mth}, and the directive-based OpenMP tasking model~\cite{omptask}.

\subsection{Dynamic Load Balancing and Work Stealing} \label{sec:dynamic-load-balancing-and-work-stealing}

Task-parallel workloads often exhibit input-dependent and irregular task costs, making \emph{static} pre-assignment of tasks to workers ineffective.
Thus, \emph{dynamic} load balancing is essential.
A widely used approach is work stealing~\cite{work-stealing}, where each worker maintains a private task deque (double-ended queue).
A worker \emph{pushes} newly created tasks to its own deque and primarily \emph{pops} from it, but \emph{steals} from another worker only when it becomes idle.
This design reduces contention compared to a centralized global queue and typically preserves locality, since a worker tends to execute tasks it recently created unless stealing occurs.

As a simple alternative, the \emph{global-queue approach} uses a single shared queue that all workers concurrently push to and pop from.
We compare the two approaches on GPUs in \secref{gpu-side-scheduler-comparison}.

\subsection{GPU Architecture} \label{sec:gpu-architecture}

NVIDIA GPUs expose a hierarchical execution model: a kernel launches a grid of \emph{thread blocks}, each consisting of \emph{warps} of 32 threads~\cite{cuda}.
A warp is the fundamental scheduling unit and executes in SIMT fashion; when threads in a warp take different control-flow paths, the paths are serialized (\emph{warp divergence}), reducing effective throughput.
Threads within a block run on the same \emph{Streaming Multiprocessor (SM)} and can cooperate through fast shared memory and barrier synchronization (e.g., \texttt{\_\_syncthreads()}).

Two properties of the memory system matter for our runtime design.
First, the per-SM L1 cache is not coherent across SMs; the L2 cache, shared by all SMs, is the coherence point, so cross-SM data must be accessed via L1-bypassing loads or atomics.
Shared memory, in contrast, is on-chip storage private to a block: data used only within one block can be kept there and accessed at low latency.
Second, CUDA provides a weakly ordered memory model, so publishing data to other threads requires explicit fences.

\section{Related Work} \label{chap:related-work}

We review prior GPU task-parallel systems against three requirements: (i) GPU-resident dynamic load balancing, (ii) fork-join with in-place parent resumption, and (iii) support for both thread and block execution modes.
As \tabref{related-work-comparison} shows, to the best of our knowledge, no prior system satisfies all three.
Here, \emph{in-place parent resumption} means that, after its children complete, the parent resumes from the same logical context with its live state preserved.

\paragraph*{Kernel-Granularity Tasking.}

Task parallelism on GPUs is well established at \emph{kernel} granularity: runtime systems such as Kokkos~\cite{kokkos3}, Legion~\cite{legion}, and Charm++~\cite{charmpp} schedule kernel-level tasks from the host.
Kokkos previously also provided a device-side task DAG interface based on futures and explicit respawning, but required programmer-managed continuation state rather than transparently resuming a suspended parent context.
CUDA Dynamic Parallelism~\cite{cuda} enables device-side kernel launches, although per-launch overhead makes it impractical for fine-grained tasks.
GTaP addresses the complementary regime of sub-kernel tasks scheduled entirely on the device.

\paragraph*{GPU-Resident Task Runtime Systems (Not Focused on Fork-Join).}

Many GPU-resident frameworks are built on the \emph{persistent-kernel} model, where a long-lived kernel repeatedly fetches and executes tasks on the device~\cite{tzeng2010}.
Representative systems include Softshell~\cite{softshell}, Whippletree~\cite{whippletree}, and Atos~\cite{atos, atos-multi-gpu}.
They provide autonomous GPU-side scheduling, sometimes at multiple execution granularities, but not fork-join with in-place parent resumption.

\paragraph*{GPU Execution Mechanisms for Fork-Join.}

Kiuchi et al.~\cite{kiuchi2025} implement fine-grained fork-join by treating program continuations as objects and repeatedly launching kernels while selecting the continuation type.
While effective for expressing resumption, this approach relies on host-driven continuation management and requires programmers to manually decompose control flow, whereas GTaP automates this transformation in the compiler.
Chatterjee et al.~\cite{chatterjee2013} describe a GPU work-stealing runtime system with finish-async style synchronization; however, their design targets block mode only and, to the best of our knowledge, no implementation is publicly available.
Tzeng et al.~\cite{tzeng2012} propose explicit dependency resolution, which can represent fork-join by modeling post-join work as dependent tasks but does not provide in-place resumption of the suspended
parent.
Zheng et al.~\cite{zheng_gpu_coroutines_2024} support in-kernel coroutine suspension for rendering, but do not provide a general fork-join task system.

\begin{table}[tb]
  \centering
  \caption{
    Comparison with prior GPU task-parallel systems along the three requirements.
    GPU-res. LB: D = distributed work redistribution, S = shared/global-queue load balancing without peer-to-peer redistribution, $\times$ = no device-side load balancing.
    Fork-join: $\checkmark$ = with in-place parent resumption, $\triangle$ = without it, $\times$ = unsupported.
    Exec.\ unit: B = block, W = warp, T = thread.
  }
  \label{tab:related-work-comparison}
  \begin{tabular}{lccc}
    \toprule
    \textbf{System} & \textbf{GPU-res.\ LB} & \textbf{Fork-join} & \textbf{Exec.\ unit} \\
    \midrule
    \begin{tabular}[c]{@{}l@{}}
      Kernel-granularity systems\\
      (Kokkos~\cite{kokkos3}, Legion~\cite{legion},\\
      Charm++~\cite{charmpp})
    \end{tabular}
           & $\times$ & \textit{N/A} & kernel \\
    Softshell~\cite{softshell}
           & S          & $\times$                  & B/T \\
    Whippletree~\cite{whippletree}
           & S          & $\times$                  & B/W/T \\
    Atos~\cite{atos, atos-multi-gpu} 
           & S          & $\times$                  & B/W/T \\
    Kiuchi et al.~\cite{kiuchi2025}
           & $\times$   & $\triangle$              & T \\
    Chatterjee et al.~\cite{chatterjee2013}
           & D          & $\triangle$              & B \\
    Tzeng et al.~\cite{tzeng2012}
           & S          & $\triangle$               & W \\
    \textbf{GTaP (ours)} 
           & \textbf{D} & $\boldsymbol{\checkmark}$ & \textbf{B/T} \\
    \bottomrule
  \end{tabular}
\end{table}

\section{Programming Model} \label{chap:programming-model}

GTaP provides a pragma-based interface for task-parallel execution on top of CUDA C++, together with a small set of runtime functions.
The API is provided by \texttt{gtap\_thread.cuh} and \texttt{gtap\_block.cuh}; including either header selects thread or block mode, respectively.

On the host, \texttt{gtap\_initialize(config)} preallocates the task-management storage, and \texttt{gtap\_finalize()} releases it.
The configuration specifies at runtime the launch geometry, preallocated task capacity, and, in thread mode, the number of DAQ queues (see \secref{api-for-thread-mode}).
Only the block size in block mode is a compile-time parameter, because join-crossing state is laid out as per-thread arrays in each task record.
\texttt{gtap\_launch(kernel, args\ldots)} launches \texttt{kernel} with the stored geometry, forwarding \texttt{args} to it.
Exceeding the capacity is detected at runtime and aborts execution with a structured error report; graceful degradation (e.g., inlining spawns under exhaustion) is future work.

\progref{adaptive-integration} presents the same adaptive numerical integration algorithm under both modes.

\begin{program*}[tb]
  \centering
  \caption{
    Adaptive numerical integration in GTaP.
  }
  \label{prog:adaptive-integration}
  \begin{minipage}[t]{0.48\textwidth}
    \subcaption{
      Thread mode.
      \texttt{trap(a,b)} approximates the integral over $[a,b]$ with a single trapezoid, sequentially on one thread.
    }
    \label{prog:adaptive-integration-thread-gtap}
    \begin{lstlisting}[keywords={}]
***#include "gtap_thread.cuh"***
__device__ double d_result;

***#pragma gtap function***
__device__ double integrate(double a, double b) {
  double m = 0.5 * (a + b);
  double coarse = trap(a, b);
  double fine   = trap(a, m) + trap(m, b);
  if (fabs(fine - coarse) < EPS)
    return (4.0 * fine - coarse) / 3.0;  // Richardson extrapolation
  double l, r;
  ***#pragma gtap task***
  l = integrate(a, m);
  ***#pragma gtap task***
  r = integrate(m, b);
  ***#pragma gtap taskwait***
  return l + r;
}
__global__ void exec_kernel(double lo, double hi) {
  ***#pragma gtap entry***
  d_result = integrate(lo, hi);
}
int main() {
  gtap_thread_config config{...};
  ***gtap_initialize(config)***;
  ***gtap_launch(exec_kernel, 0.0, 1.0)***;
  ...
}
    \end{lstlisting}
  \end{minipage}
  \hfill
  \begin{minipage}[t]{0.48\textwidth}
    \subcaption{
      Block mode.
      \texttt{block\_trap(a,b,n)} divides $[a,b]$ into \texttt{n} trapezoids evaluated in parallel by the block's threads, and broadcasts the sum.
    }
    \label{prog:adaptive-integration-block-gtap}
    \begin{lstlisting}[keywords={}]
***#include "gtap_block.cuh"***
__device__ double d_result[GTAP_BLOCK_SIZE];

***#pragma gtap function***
__device__ double integrate(double a, double b) {
  constexpr int NTRAP = 256;
  double coarse = block_trap(a, b, NTRAP);
  double fine   = block_trap(a, b, 2 * NTRAP);
  if (fabs(fine - coarse) < EPS)
    return (4.0 * fine - coarse) / 3.0;
  double m = 0.5 * (a + b), l = 0.0, r = 0.0;
  // spawn is per-thread; guard limits it to one
  ~~~if (threadIdx.x == 0)~~~ {(*@\label{line:adaptive-integration-spawn-guard}@*)
    ***#pragma gtap task***
    l = integrate(a, m);
    ***#pragma gtap task***
    r = integrate(m, b);
  }
  ***#pragma gtap taskwait***
  return l + r;  // thread 0 carries the sum
}
__global__ void exec_kernel(double lo, double hi) {
  ***#pragma gtap entry***
  d_result[threadIdx.x] = integrate(lo, hi);
}
int main() {
  gtap_block_config config{...};
  ***gtap_initialize(config)***;
  ***gtap_launch(exec_kernel, 0.0, 1.0)***;
  ...
}
    \end{lstlisting}
  \end{minipage}
\end{program*}

\subsection{Common Pragma Interface} \label{sec:common-pragma-api}

\paragraph*{\texttt{\#pragma gtap function}.}

A \texttt{\_\_device\_\_} function annotated with \texttt{\#pragma gtap function} is treated as a \emph{task function} and can be executed as a task.
The \texttt{task} and \texttt{taskwait} directives may appear only within task functions.

\paragraph*{\texttt{\#pragma gtap task}.}

A child task is spawned by placing \texttt{\#pragma gtap task} immediately before a call to a task function, optionally written as an assignment to capture its return value (e.g., \texttt{l=integrate(a,m);}).
The spawn does not wait for the child; if the call is written as an assignment, the assigned value must not be used until the corresponding \texttt{taskwait} has completed.
Task arguments are copied at spawn time (\texttt{firstprivate}-like semantics); no other data-sharing clauses are currently provided.

\paragraph*{\texttt{\#pragma gtap taskwait}.}

\texttt{taskwait} waits for the completion of all \emph{direct} child tasks spawned since the previous \texttt{taskwait} in the same task function.
A task function may contain multiple \texttt{taskwait} directives, including within conditionals and loops.
The continuation after \texttt{taskwait} is implemented by re-entry into the task function.

\paragraph*{\texttt{\#pragma gtap entry}.}

\texttt{entry} marks the root task call of a GTaP persistent kernel.
As with \texttt{task}, it must immediately precede a call to a task function, optionally written as an assignment.
\texttt{entry} is \emph{collective}: every thread of the launched grid must reach it, the root task is enqueued exactly once, and all threads then enter the persistent scheduler loop.
They leave the loop when every task has completed.

\subsection{Thread Mode} \label{sec:api-for-thread-mode}

In thread mode, each task is executed by a single CUDA thread in a sequential style.
This mode is natural for programmers and allows an easy transition from CPU task-parallel systems.

\paragraph*{The \texttt{queue} Clause}

As a scheduling hint, \texttt{\#pragma gtap task} and \texttt{\#pragma gtap taskwait} may take an optional \texttt{queue(expr)} clause.
In thread mode, the runtime system dynamically packs up to 32 tasks into a warp and executes them simultaneously, one per lane; if the packed tasks follow different control-flow paths, the warp diverges.
GTaP maintains a configurable number of task queues and fills each warp with tasks from a single queue.
By assigning tasks with similar expected control flow to the same queue, the programmer can reduce divergence among the lanes.

For \texttt{task}, \texttt{expr} selects the queue of the spawned child; for \texttt{taskwait}, it selects the queue of the post-join continuation.
We call this mechanism \emph{Divergence-Aware Queueing (DAQ)}.
The expression is evaluated at runtime, and omitting the clause is equivalent to \texttt{queue(0)}.
The clause does not change program semantics.

For example, mergesort tasks can be classified as cutoff leaves, pre-join recursive splits, or post-join merges based on the subproblem size and assigned to separate queues.
DAQ therefore applies when control-flow classes are predictable at enqueue time, not when divergence depends on branches resolved during task execution.

\subsection{Block Mode} \label{sec:api-for-block-mode}

In block mode, each task is executed cooperatively by all threads of one thread block.
A task function may therefore use block-level cooperation through \texttt{\_\_syncthreads()} and shared memory.

The spawn at \texttt{\#pragma gtap task} is performed by each thread that reaches the pragma---hence the guard at \lineref{adaptive-integration-spawn-guard} of \progref{adaptive-integration-block-gtap}, which limits each spawn to one task---while the spawned task is executed by a whole thread block.
Spawning multiple children deliberately (e.g., one child per thread) is allowed when that is the intended parallelism.
The \texttt{queue} clause is not supported: all threads of a thread block execute the same task, so tasks are never intermixed within a warp.
\texttt{\#pragma gtap taskwait} must be reached by all threads of the block, in the same sense as \texttt{\_\_syncthreads()}.

\paragraph*{Values across spawn and join.}

A block-mode task call behaves like an ordinary block-wide device-function call, made asynchronous.
Arguments are broadcast: the child's threads all observe the values evaluated by the spawning thread.
Local variables retain their per-thread values across the join, even when the continuation resumes on a different thread block.
Return values are per-thread (thread $i$ observes the child's thread $i$) and are assigned only on the thread that executed the spawn.
Thus, in \progref{adaptive-integration-block-gtap}, only thread~0 receives \texttt{l} and \texttt{r}, while the others retain \texttt{0.0}.

\subsection{Restrictions} \label{sec:restrictions}

GTaP's re-entry mechanism imposes two semantic restrictions.
Values that cross \texttt{taskwait} must be trivially copyable.
A continuation is not guaranteed to resume on the same thread (thread mode) or block (block mode), because the re-enqueued task may be stolen; state tied to the physical execution context (e.g., \texttt{threadIdx}, shared memory) must not be carried across \texttt{taskwait}.

\section{Implementation} \label{chap:implementation}

\subsection{GTaP Runtime System} \label{sec:gtap-runtime}

\subsubsection{Overview} \label{sec:overview-of-runtime-design-and-implementation}

GTaP executes tasks within a persistent kernel: all threads of the launched grid enter a scheduler loop that repeatedly acquires and executes tasks entirely on the GPU.
The scheduling unit is a warp in thread mode and a thread block in block mode.
Each task has a \emph{task record} containing its payload, scheduling metadata, parent/child information, and resumption state.
Each task is identified by a \emph{task ID}, an index into the storage holding its record.
In thread mode each warp owns a local deque of runnable task IDs; in block mode each block owns one.
The task-record storage and task queues are preallocated in GPU memory before execution to avoid device-side dynamic allocation during kernel execution.

\algref{thread-mode-scheduler} shows the persistent scheduler loop executed by each warp in thread mode.
Batching the publication in \textsc{PushBatch} pays the memory fence and the atomic queue-size update once per batch rather than once for each task.

Though we omit the details, block mode uses an analogous scheduler loop per block: it acquires at most one task by single-task pop or steal, executes it cooperatively with all threads, and publishes newly runnable tasks to its local deque, synchronizing block-wide where required.

\begin{algorithm}[tb]
  \caption{
    Persistent scheduler loop executed by each warp in thread mode.
    DAQ is omitted for simplicity.
    Runnable tasks occupy the lowest $\mathit{cnt}$ lanes.
    \textsc{Retain} packs retained tasks from lane~0, while \textsc{PopBatch} and \textsc{StealBatch} fill the following lanes.
    \textsc{PopBatch}, \textsc{StealBatch}, and \textsc{PushBatch} are warp-collective queue operations; \textsc{PopBatch} is detailed in \algref{pop-batch}.
  }
  \label{alg:thread-mode-scheduler}
  \Fn{\textsc{ThreadModeScheduler}()}{
    \Comment{per-lane: $\mathit{tid}$ (this lane's task ID, valid only when $\LANE<\mathit{cnt}$)}
    \Comment{warp-uniform: $\mathit{cnt}$ (number of lanes with a task)}
    \Comment{warp-shared: $\mathit{staged}$ (task IDs made runnable this iteration)}
    $\mathit{cnt} \leftarrow 0$\;

    \lIf{\textsc{IsGridLeader}()}{\textsc{EnqueueRootTask}()}

    \While{\textbf{true}}{
      \Comment{lanes $[0,\,\mathit{cnt})$ hold retained tasks}
      \If{$\mathit{cnt} < 32$}{
        $\mathit{popped} \leftarrow \textsc{PopBatch}(Q_{\mathrm{self}},\,\mathit{cnt},\,\mathit{tid})$\;
        $\mathit{cnt} \leftarrow \mathit{cnt} + \mathit{popped}$\;
      }
      \If{$\mathit{cnt} < 32$}{
        $Q_{\mathrm{victim}} \leftarrow \textsc{SelectVictimQueue}()$\;
        $\mathit{stolen} \leftarrow \textsc{StealBatch}(Q_{\mathrm{victim}},\,\mathit{cnt},\,\mathit{tid})$\;
        $\mathit{cnt} \leftarrow \mathit{cnt} + \mathit{stolen}$\;
      }

      \If{$\mathit{cnt} = 0$}{
        \lIf{\textsc{AllTasksComplete}()}{\Return}
        \KwCont\;
      }

      $\mathit{staged} \leftarrow \emptyset$\;
      \If{$\LANE < \mathit{cnt}$}{
        $\mathit{staged} \mathrel{{+}{=}}$ execute task $\mathit{tid}$, collecting newly runnable task IDs\;
      }
      \SyncWarp{}\;

      \Comment{retain up to 32 staged IDs}
      $(\mathit{tid},\,\mathit{cnt},\,\mathit{rest}) \leftarrow \textsc{Retain}(\mathit{staged},\,32)$\;
      \Comment{publish the rest to the local queue}
      $\textsc{PushBatch}(Q_{\mathrm{self}},\,\mathit{rest})$\;
    }
  }
\end{algorithm}

\subsubsection{Implementation of Fork-Join} \label{sec:fork-join-implementation}

When a task reaches a join, the runtime stores its live state and resumption point in its task record and returns to the scheduler.
A waiting-child counter records the number of unfinished children.
As each child completes, it writes its results, executes a memory fence, and atomically decrements this counter.
The child that decrements the counter to zero re-enqueues the parent, whose next invocation resumes from the recorded point.
The fence ensures the resumed parent observes all children's results.
Because resumption is an ordinary re-enqueue, the continuation can be routed like any newly spawned task---this is why \texttt{taskwait} accepts a \texttt{queue} clause (\secref{api-for-thread-mode}).

\subsubsection{Work Stealing and Task Queue} \label{sec:work-stealing-and-task-queue-implementation}

GTaP uses GPU-resident random work stealing: the owner of each deque pops from the tail, and idle thieves steal from the head.

\paragraph*{Block Mode.}

For block mode, we place one deque per block.
Each block designates a leader thread for queue operations; each pop/steal retrieves at most one task, and a retrieved task is executed by all threads of the block.
The design is based on the Chase--Lev work-stealing deque~\cite{chase-and-lev-deque}, which provides a fast lock-free path for owner operations (push/pop).

\paragraph*{Thread Mode.}

For thread mode, we place one deque per warp (multiple deques when DAQ is enabled).
In each persistent-kernel iteration, a warp acquires up to 32 runnable tasks by a batched pop/steal---a warp-cooperative operation detailed below---and executes them, one task per lane.
Pushes are batched as well: the warp keeps up to 32 newly runnable tasks for immediate execution and enqueues the rest.
With DAQ enabled, each pop or steal first selects the fullest deque---the one holding the most available tasks---and tries other deques if necessary.

\emph{Deque representation.}
Each deque is a fixed-capacity ring buffer allocated at initialization, with logical pointers \texttt{head} and \texttt{tail}: \texttt{head} is the steal end and \texttt{tail} is the owner end.
We additionally maintain \texttt{count}, the number of \emph{available (not-yet-claimed)} tasks.
\texttt{head} and \texttt{count} reside in global memory and are accessed with L1-bypassing loads and stores through L2, the coherence point across SMs; in our implementation we use PTX cache operators (\texttt{ld.global.cg} and \texttt{st.global.cg}).
\texttt{tail} is kept in shared memory: only the owner warp updates it, and thieves never access it---the number of stealable entries is conveyed solely through \texttt{count}.
A per-queue \texttt{lock} serializes steals so that at most one thief steals from a victim at a time.

\emph{Batched pop.}
\algref{pop-batch} shows \textsc{PopBatch}.
Lane~0 atomically \emph{claims} up to 32 tasks by decrementing \texttt{count} via CAS and advances \texttt{tail} locally; the claimed size is then broadcast to the warp, and lanes load the corresponding task IDs from the tail end in parallel.

\begin{algorithm}[tb]
  \caption{
    \textsc{PopBatch}: warp-cooperative batched pop from a warp-local deque.
  }
  \label{alg:pop-batch}

  \Fn{\textsc{PopBatch}($Q,\,\mathit{cnt},\,\mathit{tid}$)}{
    \KwIn{$Q$: warp-local queue}
    \KwIn{$\mathit{cnt}$: number of lanes already holding retained tasks; at most $32 - \mathit{cnt}$ tasks are popped}
    \KwInOut{$\mathit{tid}$: this lane's task ID; written only on lanes $[\mathit{cnt},\,\mathit{cnt} + \mathit{pop\_count})$}
    \KwOut{$\mathit{pop\_count}$: number of tasks popped}
    \Comment{$\mathit{tail}$: $Q$'s logical tail, kept in shared memory across iterations; updated only by the owner warp}

    \If{$\LANE = 0$}{
      \While{\textbf{true}}{
        $\mathit{old\_q\_cnt} \leftarrow$ \Load{$Q.\mathit{count}$}\;
        \If{$\mathit{old\_q\_cnt} \le 0$}{
          $\mathit{pop\_count} \leftarrow 0$\;
          \KwBreak\;
        }
        $\mathit{claim} \leftarrow$ \MIN{$32 - \mathit{cnt}$, $\mathit{old\_q\_cnt}$}\;
        \If{\CAS{$Q.\mathit{count}$, $\mathit{old\_q\_cnt}$, $\mathit{old\_q\_cnt} - \mathit{claim}$}
            $= \mathit{old\_q\_cnt}$}{
          $\mathit{pop\_count} \leftarrow \mathit{claim}$\;
          $\mathit{tail} \leftarrow \mathit{tail} - \mathit{claim}$\;
          \KwBreak\;
        }
      }
    }

    \Comment{\SHFL{} also synchronizes the warp}
    $\mathit{pop\_count} \leftarrow$ \SHFL{$\mathit{pop\_count}$, $0$}\;

    \If{$\LANE \in [\mathit{cnt},\,\mathit{cnt} + \mathit{pop\_count})$}{
      $\mathit{offset} \leftarrow \LANE - \mathit{cnt}$\;
      $\mathit{idx} \leftarrow (\mathit{tail} + \mathit{offset}) \bmod \operatorname{capacity}(Q)$\;
      $\mathit{tid} \leftarrow$ \Load{$Q.\mathit{queue}[\mathit{idx}]$}\;
    }

    \Return $\mathit{pop\_count}$\;
  }
\end{algorithm}

\emph{Batched steal and push.}
\textsc{StealBatch} mirrors \textsc{PopBatch} on the head end.
A thief locks the victim deque, claims tasks by CAS on \texttt{count}, loads the claimed IDs, and only then advances \texttt{head} and releases the lock.
\textsc{PushBatch} first executes \texttt{\_\_threadfence()}, then publishes enqueued task IDs by incrementing the deque's \texttt{count}.

\emph{Correctness and memory ordering.}
Safety relies on three properties.
First, \texttt{count} tracks unclaimed entries, and every claim must succeed in a CAS on \texttt{count}.
Concurrent claims are thereby serialized, so each queue entry is claimed at most once and the owner and thieves cannot claim overlapping regions of the ring buffer.
They then advance opposite ends only within their claimed regions, while the victim lock serializes steals; thus, each claimed queue entry is fetched at most once.
Second, stores of enqueued task IDs are ordered by a fence before \texttt{count} is incremented, so a consumer that subsequently claims them observes initialized queue entries.
Finally, claimed-but-unread slots are protected across wrap-around: a thief advances \texttt{head} only after loading the task IDs from its claimed slots, and the owner's capacity check is based on \texttt{head} and \texttt{tail} rather than \texttt{count}, so those slots are not prematurely overwritten.

\subsection{GTaP Compiler} \label{sec:gtap-compiler}

We extend Clang to lower each function annotated with \texttt{\#pragma gtap function} into a suspendable, resumable function by rewriting the CUDA device AST.
The compiler generates a task-record layout for the values that must survive suspension---arguments and locals that cross a \texttt{taskwait}.
For block mode, spilled values are laid out as per-thread arrays, realizing the per-thread value semantics of \secref{api-for-block-mode}.
Each \texttt{task} directive is lowered to runtime operations that create a child task.
At a \texttt{taskwait}, the transformed function records its resumption state and returns to the scheduler.
To resume the task, the runtime invokes the function again, and a generated \texttt{switch} transfers control to the recorded continuation point.

\section{Performance Evaluation} \label{chap:evaluation}

We evaluate GTaP on one NVIDIA GH200 node of the Miyabi-G supercomputer~\cite{miyabi}, comprising a 72-core Grace CPU (512\,GB/s) and a single H100 GPU (96\,GB HBM3, 4.0\,TB/s).
GTaP is compiled with our Clang~21.1.8 extension using CUDA Toolkit~12.9 and \texttt{-O3 -x cuda -{}-cuda-gpu-arch=sm\_90}; CPU implementations use \texttt{-O3} on the Grace CPU.

We report the median over 20 runs.
For GTaP, we measure the execution phase only, excluding one-time initialization and result retrieval; the two exceptions are the end-to-end FMM breakdown and the $k$-clique comparison, which include GTaP's initialization as part of the measured pipeline.
For CPU executions, we exclude runtime startup by warming up with a dummy parallel region before timing.

\subsection{Understanding the Execution Modes}
\label{sec:understanding-execution-modes}

First, we compare the two execution modes of GTaP through a microbenchmark.
For tasks without intra-task parallelism, thread mode naturally exploits parallelism across tasks.
When a task contains data-parallel work, both modes can be effective: block mode exploits parallelism within each task, whereas thread mode runs multiple sequentially written tasks in each warp.
We study this on a full binary task tree whose leaves each run the same data-parallel compute loop: the loop count (\texttt{compute\_iters}) sets the task grain, and the tree depth ($D$) sets the available parallelism.
Both modes use grid size 2000 and block size 32.

\paragraph*{Idealized Cost Model.}
For a leaf task, let $t_{\mathrm{th}}$ and $t_{\mathrm{bl}}$ denote its execution times in thread and block modes, respectively ($t_{\mathrm{bl}}\ge t_{\mathrm{th}}/32$).
Let $s$ denote the per-step scheduling cost, which is comparable between the two modes because both execute the same leader-side queue operations.
With enough runnable tasks to fill every lane, one scheduler step processes 32 leaves in thread mode but only one in block mode, so the leaf-processing times relate as
$T^{\mathrm{leaf}}_{\mathrm{block}} / T^{\mathrm{leaf}}_{\mathrm{thread}}
\approx 32\,(t_{\mathrm{bl}}+s)/(t_{\mathrm{th}}+s) 
\ge (t_{\mathrm{th}}+32s)/(t_{\mathrm{th}}+s)
>1$.
This steady-state model isolates leaf processing and omits internal-node overhead, such as spawn and join operations.

Block mode can reduce this scheduling overhead through a cutoff: instead of spawning tasks down to individual leaves, it executes a subtree as one block task.
A 5-level cutoff groups 32 leaves into one task, recovering the batching that thread mode obtains automatically.
This improves performance, but requires programmers to manually tune task granularity.

\begin{figure}[tb]
  \centering
  \includegraphics[width=\linewidth]{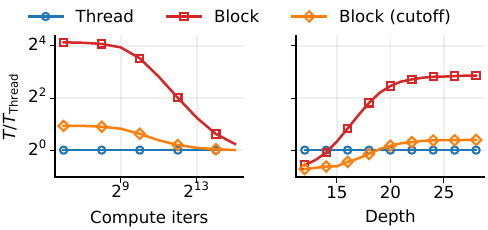}
  \caption{
    Homogeneous full binary tree: execution time normalized to thread
    mode.
    Left: $\texttt{compute\_iters}$ sweep ($D=25$);
    right: depth~$D$ sweep ($\texttt{compute\_iters}=2048$).
    Block~(cutoff) stops spawning 5 levels above leaves.
  }
  \label{fig:mode-tree-ratio}
\end{figure}

\paragraph*{Results.}
\figref{mode-tree-ratio} shows the two complementary regimes.
In the granularity sweep (left), plain block mode starts
$17\times$ slower because it pays scheduling overhead for each task,
whereas thread mode amortizes the same overhead across a warp batch.
As leaf computation grows (increasing both $t_{\mathrm{th}}$ and $t_{\mathrm{bl}}$ relative to $s$), scheduling overhead becomes less significant and the two modes converge.
The cutoff variant substantially narrows the gap by manually
coarsening block-mode tasks.

The depth sweep (right) exposes the effect of available task
parallelism.
For shallow trees, the runnable frontier is too small to fill all warp
lanes, making block mode preferable.
As the tree becomes deeper, more leaves become simultaneously
runnable, allowing thread mode to keep warps fully occupied and process
32 independent tasks per scheduling step.
Plain block mode leads for $D\le14$, and the cutoff variant for
$D\le18$; on deeper trees, thread mode is faster.

Thus, thread mode is preferable for abundant homogeneous tasks even when each task contains data-parallel work, while block mode is useful when task-level parallelism is limited.

\subsection{Effect of Divergence-Aware Queueing}
\label{sec:effect-of-daq}

A major enemy of thread mode is inter-task warp divergence.
The analysis above assumes homogeneous leaf tasks; we introduce heterogeneity by assigning each leaf one of $K$ equal-cost compute functions with distinct control-flow paths.
The function is selected by hashing the node ID and is therefore known at spawn time.

Block mode is unaffected because each block executes a single task.
Thread mode, however, suffers when a warp batch contains multiple task types.

\begin{figure}[tb]
  \centering
  \includegraphics[width=\linewidth]{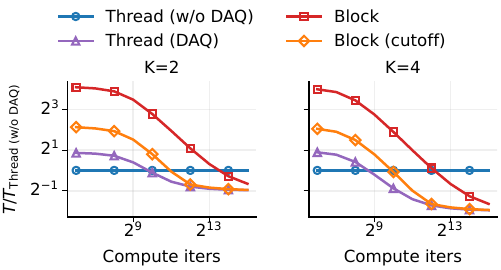}
  \caption{
    Heterogeneous-leaf binary tree: execution time normalized to thread mode without DAQ.
    Left: $K=2$; right: $K=4$.
    Both panels sweep $\texttt{compute\_iters}$ at $D=25$.
  }
  \label{fig:hetero-normalized}
\end{figure}

DAQ addresses this by allowing programmers to specify a queue type.
\figref{hetero-normalized} shows execution time normalized to thread
mode without DAQ ($K=2$ and $4$).
Thread mode without DAQ is up to $\sim K\times$ slower than block mode due to warp divergence.
With DAQ, all tasks in a warp batch execute the same compute function, eliminating this divergence-induced slowdown.
Thus, DAQ preserves the programmability of fine-grained thread mode while recovering the efficiency of homogeneous task execution.

\subsection{Microbenchmarks} \label{sec:microbenchmarks}

We conduct two experiments on three recursive task-parallel workloads.
First, with DAQ disabled, we compare GTaP with existing GPU and CPU runtimes by sweeping the problem size under fixed cutoffs.
Second, we evaluate DAQ separately by fixing the problem size and sweeping the cutoff.
All three workloads use thread mode because each task body is naturally sequential.

\paragraph*{Workloads.}

We use three workloads that stress different aspects of GPU-resident fork-join execution.
\emph{Fibonacci} represents extremely fine-grained recursion and primarily stresses task-management overheads.
\emph{N-Queens} represents highly irregular task generation due to pruning; we count solutions via bitmask-based backtracking.
\emph{Cilksort} parallelizes merging to remove mergesort's sequential tail.

\subsubsection{Runtime Comparison} \label{sec:microbenchmark-runtime-comparison}

\paragraph*{Baselines and Setup.}

To the best of our knowledge, the continuation-based runtime system of Kiuchi et al.~\cite{kiuchi2025} is the only prior implementation of general-purpose fine-grained fork-join task parallelism on GPUs.
We port their implementation to GH200 with minor fixes and compile it with \texttt{nvcc} from the same CUDA toolkit at \texttt{-O3} for the same target.\footnote{The modifications are included in the artifact.}

As supplementary reference points, we use LLVM OpenMP tasks and OpenCilk~\cite{opencilk} on the 72-core Grace CPU.
These cross-architecture numbers position GPU-resident fork-join against mature CPU runtime systems rather than compare the runtime systems head-to-head.

For this comparison, Fibonacci disables the cutoff and spawns a task at every recursive call; N-Queens uses a fixed cutoff depth of 7; and Cilksort uses sort/merge cutoffs tuned at $n=10^9$: 64/256 on the GPU systems and 4096/4096 on the CPU systems.
For each benchmark, we tune the grid/block configuration at the largest measured problem size and hold it fixed while sweeping the problem size.

\paragraph*{Results.}

\begin{figure*}[tb]
  \centering
  \includegraphics[width=0.9\linewidth]{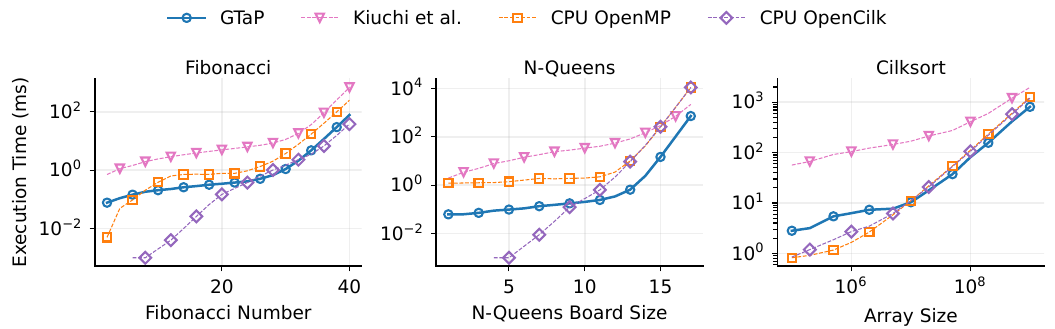}
  \caption{
    Microbenchmark execution time (medians over 20 runs) across problem sizes.
    Lower is better.
  }
  \label{fig:microbenchmark-performance-comparison}
\end{figure*}

\figref{microbenchmark-performance-comparison} plots the full sweep; \tabref{microbenchmark-relative-time} reports GTaP's speedup over each baseline at the largest size.

\begin{table}[tb]
  \centering
  \caption{
    Speedup of GTaP over each baseline at the largest problem size for each benchmark.
    Values report $T_{\mathrm{baseline}}/T_{\mathrm{GTaP}}$; above 1 indicates that GTaP is faster.
  }
  \label{tab:microbenchmark-relative-time}
  \small
  \begin{tabular}{lrrrr}
    \toprule
    \textbf{Benchmark} & \textbf{Size} & \textbf{Kiuchi et al.} & \textbf{OpenMP} & \textbf{OpenCilk} \\
    \midrule
    Fibonacci & $n=40$     & 8.8 & 3.2 & 0.49 \\
    N-Queens  & $n=17$     & 3.0 & 15.1 & 15.3 \\
    Cilksort  & $n=10^9$   & 2.4 & 1.6 & 1.5 \\
    \bottomrule
  \end{tabular}
\end{table}

\textbf{Fibonacci:}
GTaP is $8$--$15\times$ faster than the GPU baseline: the baseline drives execution through repeated kernel launches from the host and manages continuation objects in GPU memory, whereas GTaP schedules and resumes tasks entirely inside one persistent kernel with preallocated task records.
At $n=40$, GTaP is $3.2\times$ faster than OpenMP and only $2.0\times$ slower than OpenCilk, despite the negligible work per leaf.

\textbf{N-Queens:}
GTaP is over $100\times$ faster than the GPU baseline where task management dominates ($n\approx8$--$12$).
It also outperforms both CPU runtime systems by ${\sim}15\times$: the leaf work is register- and bitwise-heavy and suits GPU execution.

\textbf{Cilksort:}
GTaP overtakes both CPU runtime systems for $n\ge10^7$, though the gap narrows to ${\sim}1.5\times$ at $n=10^9$ as memory bandwidth becomes the bottleneck.
GTaP's modest speedup over the CPU baselines---despite an $8\times$ raw-bandwidth advantage---reflects a structural cost of thread-mode task parallelism: lanes of a warp execute unrelated tasks on unrelated data, so accesses are rarely coalesced and the GPU realizes only a fraction of its peak bandwidth.

\subsubsection{DAQ Applicability} \label{sec:microbenchmark-daq}

In a separate experiment, we fix the problem size (Fibonacci: $n=40$; N-Queens: $n=16$; Cilksort: $n=5\times10^7$) and sweep the cutoff to vary both the number of spawned tasks and the work performed by cutoff leaves.
The DAQ configuration uses three queues for Fibonacci (recursive tasks, serial-cutoff leaves, and post-\texttt{taskwait} continuations), two for N-Queens (recursive and cutoff tasks), and three for Cilksort (recursive, serial-sort, and serial-merge tasks).

\begin{figure}[tb]
  \centering
  \includegraphics[width=\linewidth]{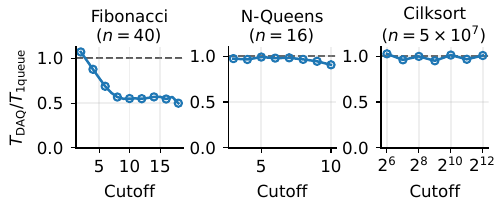}
  \caption{
    Effect of DAQ in the microbenchmarks, normalized to the single-queue
    configuration. 
    Lower is better.
  }
  \label{fig:daq-microbenchmarks}
\end{figure}

As \figref{daq-microbenchmarks} shows, DAQ improves Fibonacci by up to $2.0\times$ because the leaf bodies are internally homogeneous.
For N-Queens and Cilksort, however, substantial control-flow divergence remains within the leaf bodies themselves, so separating leaves alone provides little benefit.

\subsection{Real-World Applications} \label{sec:real-world-applications}

Beyond microbenchmarks, we apply GTaP to two workloads with irregular phases that are naturally task-parallel.

\subsubsection{FMM Dual-Tree Traversal} \label{sec:fmm-application}

We evaluate a 3D Laplace FMM pipeline with $N$ particles and multipole acceptance criterion (MAC) parameter $\theta$: adaptive octree construction, upward pass, interaction-list construction via dual-tree traversal (DTT), and data-parallel multipole-to-local (M2L) and particle-to-particle (P2P) / local-to-particle (L2P) evaluation.
DTT is naturally expressed as task-parallel recursion~\cite{dehnen2002, taura2012fmm, yokota2013dtt}, but remains difficult to execute efficiently on GPUs.
For example, Wilson et al.~\cite{wilson2021bldtt} perform DTT on the CPU, whereas Gumerov and Duraiswami~\cite{gumerov2008gpu} use stencil-defined interaction lists rather than DTT.
We instead run DTT itself as GPU-resident tasks.

The DTT phase in our pipeline only constructs interaction lists; numerical evaluation is regular and naturally handled by subsequent data-parallel GPU kernels.
Starting from the root--root pair, it applies the MAC to each cell pair, appending well-separated pairs to M2L lists and leaf--leaf pairs to P2P lists, and otherwise splitting the larger cell into child pairs.
To avoid device-side dynamic allocation during traversal, interaction lists are fixed-capacity arrays appended via \texttt{atomicAdd}; we verify capacities against overflow in every run.
Only DTT differs across three variants---(i)~GTaP task traversal (thread mode), (ii)~a level-synchronous GPU worklist, and (iii)~Host OpenMP tasks followed by a host-to-device copy of the lists---while other kernels are shared and implemented as GPU kernels.
The GPU worklist baseline uses two fixed-capacity ping-pong frontiers and atomically appends unresolved child pairs, requiring one kernel launch and host synchronization per frontier iteration.

\begin{figure}[tb]
  \centering
  \includegraphics[width=\linewidth]{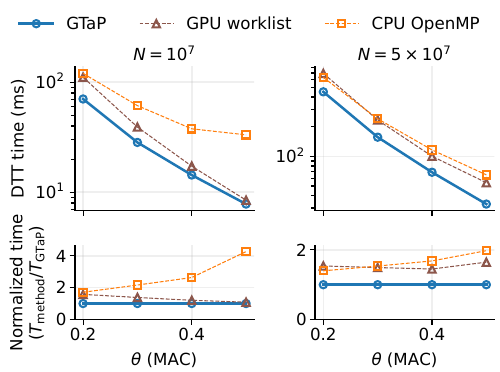}
  \caption{
    DTT list-construction time vs.\ $\theta$ for $N=10^7$ and $N=5\times10^7$ (top: absolute time; bottom: normalized to GTaP).
  }
  \label{fig:fmm-dtt-theta}
\end{figure}

\figref{fmm-dtt-theta} isolates DTT time for uniformly distributed particles across a $\theta$ sweep for $N=10^7$ and $5\times10^7$; smaller $\theta$ tightens the MAC, deepening traversal and growing the lists, thereby producing the DTT-heavy regime.
GTaP is consistently the fastest: at $N=5\times10^7$, $\theta=0.2$, it is $1.5\times$ faster than the GPU worklist and $1.4\times$ faster than Host OpenMP.

\begin{figure}[tb]
  \centering
  \includegraphics[width=\linewidth]{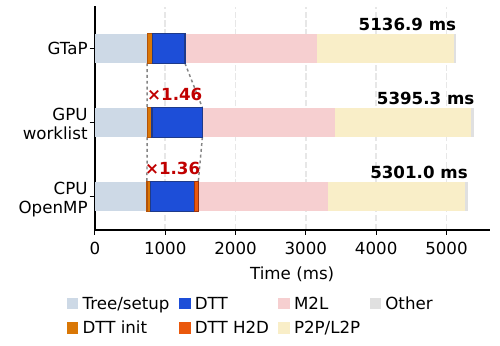}
  \caption{
    End-to-end FMM3D runtime breakdown for $N=5\times10^7$ and $\theta=0.2$ (DTT-heavy regime).
    Each bar shows an independently measured end-to-end run.
    Red labels show speedup of the DTT segment only.
  }
  \label{fig:fmm-e2e-breakdown}
\end{figure}

\figref{fmm-e2e-breakdown} shows the end-to-end breakdown at $N=5\times10^7$, $\theta=0.2$.
Main phases outside DTT---upward pass, M2L, and P2P/L2P evaluation---are data-parallel CUDA kernels shared across the three variants.
GTaP substantially shrinks the DTT segment, but M2L and evaluation dominate the pipeline, so the end-to-end speedup is limited to ${\sim}1.05\times$ over the GPU worklist and ${\sim}1.03\times$ over Host OpenMP.

\subsubsection{$k$-Clique Counting} \label{sec:k-clique-application}

We evaluate $k$-clique counting on SNAP graphs (com-DBLP, as-Skitter, and com-Orkut)~\cite{snapnets} against \textsc{KCGPU}~\cite{almasri2022}, a state-of-the-art specialized GPU counter.
\textsc{KCGPU} implements two algorithms---\emph{graph orientation} and \emph{pivoting}---with hand-tuned binary-encoded set intersection, and statically assigns the search tree of each oriented edge to one thread block.
This static edge-to-block mapping is efficient when per-edge search trees are uniform, but on skewed graphs, a few expensive searches remain assigned to individual blocks, causing load imbalance and a long execution tail.

We implement GTaP versions of \textsc{KCGPU}'s orientation and pivoting algorithms.
We use GTaP's thread mode, where each thread explores one edge.
A task splits when a cheap spawn-time estimate of its subtree cost exceeds a threshold---for orientation, the size of the edge's common out-neighbor set $C$; for pivoting, pivot-list statistics during the search.
To split, it encodes the candidate subgraph as a bitset and partitions the current branch range into child tasks; each child re-applies the same test, spawning further tasks for heavy subtrees and finishing light ones by plain recursion.
GTaP makes such dynamic splitting practical: programmers expose recursive subtasks with simple pragmas, and the runtime system redistributes them through work stealing.

\tabref{k579-count-phase-time} reports end-to-end count times, including orientation and GTaP's runtime initialization (counts verified against \textsc{KCGPU}).
Two trends emerge.
For shallow searches (small $k$, or pivoting, which bounds recursion depth), \textsc{KCGPU}'s static mapping suffices and its lower per-task overhead wins---up to $4.5\times$ on com-Orkut (orientation, $k=5$).
As the search deepens and skew increases, work stealing becomes more effective: at $k=9$, GTaP outperforms \textsc{KCGPU} by $5.2\times$ on com-DBLP and $2.3\times$ on as-Skitter with orientation, and by $1.9\times$ on as-Skitter and $1.5\times$ on com-Orkut with pivoting.

\begin{figure}[tb]
  \centering
  \begin{subfigure}[t]{\linewidth}
    \centering
    \includegraphics[width=0.95\linewidth]{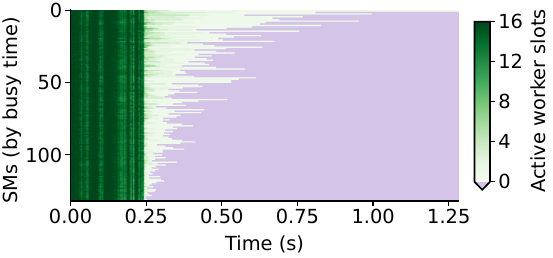}
    \subcaption{
      \textsc{KCGPU}.
      Rows are SMs; color shows the number of active block slots per SM (lavender = idle).
    }
    \label{fig:kclique-timelines-kcgpu}
  \end{subfigure}
  \vspace{0.3\baselineskip}
  \begin{subfigure}[t]{\linewidth}
    \centering
    \includegraphics[width=0.95\linewidth]{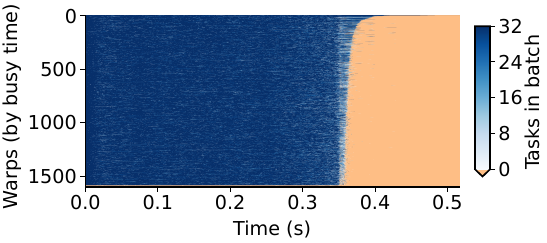}
    \subcaption{
      GTaP.
      Rows are warps in thread mode; color shows tasks per warp batch (orange = idle).
    }
    \label{fig:kclique-timelines-gtap}
  \end{subfigure}
  \caption{
    Working timelines on as-Skitter, $k=9$ (pivoting).
    Rows are sorted by total busy time.
    Timelines show count-phase kernel execution only.
  }
  \label{fig:kclique-timelines}
\end{figure}

\figref{kclique-timelines} visualizes the source of GTaP's pivoting speedup on as-Skitter at $k=9$.
In \textsc{KCGPU}, most SMs become idle while a few blocks continue processing expensive per-edge search trees.
GTaP shortens this execution tail by splitting and redistributing heavy subtrees across threads.

\begin{table}[tb]
  \centering
  \caption{
    Execution time of GTaP and \textsc{KCGPU} $k$-clique counting implementations.
  }
  \label{tab:k579-count-phase-time}
  \small
  \begin{tabular}{l|c|rr|rr|}
    \cline{3-6}
    \multicolumn{1}{l}{} & \multicolumn{1}{c|}{} & \multicolumn{4}{c|}{Execution Time (s)} \\ \hline
    \multicolumn{1}{|c|}{\multirow{2}{*}{Graph}} & \multicolumn{1}{c|}{\multirow{2}{*}{$k$}} &
    \multicolumn{2}{c|}{Graph Orientation} &    \multicolumn{2}{c|}{Pivoting} \\ \cline{3-6}
    \multicolumn{1}{|l|}{} & \multicolumn{1}{c|}{} &    \multicolumn{1}{c|}{GTaP} & \multicolumn{1}{c|}{\textsc{KCGPU}} &    \multicolumn{1}{c|}{GTaP} & \multicolumn{1}{c|}{\textsc{KCGPU}} \\ \hline
    \multicolumn{1}{|l|}{\multirow{3}{*}{com-DBLP}} & \multicolumn{1}{c|}{5} &    \multicolumn{1}{r|}{0.040} & \multicolumn{1}{r|}{\textbf{0.010}} &    \multicolumn{1}{r|}{0.161} & \multicolumn{1}{r|}{\textbf{0.116}} \\ \cline{2-6}
    \multicolumn{1}{|l|}{} & \multicolumn{1}{c|}{7} &    \multicolumn{1}{r|}{\textbf{0.271}} & \multicolumn{1}{r|}{0.317} &    \multicolumn{1}{r|}{0.159} & \multicolumn{1}{r|}{\textbf{0.115}} \\ \cline{2-6}
    \multicolumn{1}{|l|}{} & \multicolumn{1}{c|}{9} &    \multicolumn{1}{r|}{\textbf{22.287}} & \multicolumn{1}{r|}{116.112} &    \multicolumn{1}{r|}{0.158} & \multicolumn{1}{r|}{\textbf{0.114}} \\ \hline
    \multicolumn{1}{|l|}{\multirow{3}{*}{as-Skitter}} & \multicolumn{1}{c|}{5} &    \multicolumn{1}{r|}{0.131} & \multicolumn{1}{r|}{\textbf{0.058}} &    \multicolumn{1}{r|}{0.965} & \multicolumn{1}{r|}{\textbf{0.467}} \\ \cline{2-6}
    \multicolumn{1}{|l|}{} & \multicolumn{1}{c|}{7} &    \multicolumn{1}{r|}{1.011} & \multicolumn{1}{r|}{\textbf{0.541}} &    \multicolumn{1}{r|}{\textbf{0.795}} & \multicolumn{1}{r|}{0.961} \\ \cline{2-6}
    \multicolumn{1}{|l|}{} & \multicolumn{1}{c|}{9} &    \multicolumn{1}{r|}{\textbf{30.564}} & \multicolumn{1}{r|}{71.436} &    \multicolumn{1}{r|}{\textbf{0.785}} & \multicolumn{1}{r|}{1.484} \\ \hline
    \multicolumn{1}{|l|}{\multirow{3}{*}{com-Orkut}} & \multicolumn{1}{c|}{5} &    \multicolumn{1}{r|}{3.538} & \multicolumn{1}{r|}{\textbf{0.781}} &    \multicolumn{1}{r|}{10.259} & \multicolumn{1}{r|}{\textbf{6.666}} \\ \cline{2-6}
    \multicolumn{1}{|l|}{} & \multicolumn{1}{c|}{7} &    \multicolumn{1}{r|}{14.383} & \multicolumn{1}{r|}{\textbf{4.185}} &    \multicolumn{1}{r|}{12.191} & \multicolumn{1}{r|}{\textbf{10.777}} \\ \cline{2-6}
    \multicolumn{1}{|l|}{} & \multicolumn{1}{c|}{9} &    \multicolumn{1}{r|}{267.425} & \multicolumn{1}{r|}{\textbf{90.140}} &    \multicolumn{1}{r|}{\textbf{13.470}} & \multicolumn{1}{r|}{20.195} \\ \hline
  \end{tabular}
\end{table}

\subsection{GPU-Side Scheduler Comparison}
\label{sec:gpu-side-scheduler-comparison}

We compare three thread-mode scheduler designs: a centralized global queue, non-batched work stealing, and GTaP's warp-cooperative batched work stealing.
All three variants share GTaP's runtime system and differ only in queue management.
We use Fibonacci as a scheduler-bound stress test: tasks are tiny and frequently spawn children, so task execution does not hide queue-management overhead.
We sweep the number of threads by varying the grid size.

\begin{figure}[tb]
  \centering
  \includegraphics[width=\linewidth]{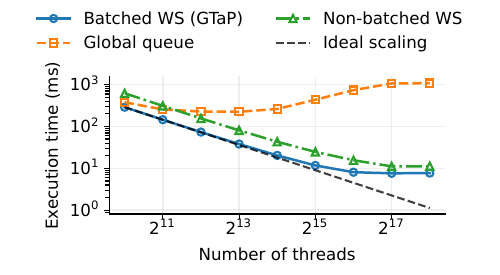}
  \caption{
    Scheduler comparison in thread mode on Fibonacci ($n=35$).
    Ideal scaling extrapolates $1/P$ ($P$: thread count) from the smallest thread count.
  }
  \label{fig:gpu-side-scheduler-comparison}
\end{figure}

\figref{gpu-side-scheduler-comparison} reports the results.
Work stealing outperforms a centralized global queue, which quickly becomes a scalability bottleneck.
Among the work-stealing variants, our warp-cooperative batched pop/steal outperforms a non-batched variant in which each warp issues up to 32 single-task pop/steal operations.
These results support GTaP's scheduler choices.

\section{Conclusion} \label{chap:conclusion}

We presented GTaP, a GPU-resident task-parallel system that supports fork-join entirely within a persistent GPU kernel through a pragma-based API.
GTaP supports block and thread modes, balances load via warp-cooperative batched work stealing, and optionally reduces warp divergence with DAQ.
On real-world irregular workloads, pragma-annotated recursion is competitive with hand-written GPU baselines and outperforms a state-of-the-art specialized kernel by up to $5.2\times$ on deep, skewed searches, showing that GPU-resident fork-join is both practical and competitive.

As future work, natural extensions include relaxing current API restrictions, richer tasking constructs (e.g., \texttt{depend}), locality-aware stealing~\cite{hotslaw, laws}, and multi-GPU support.
A further direction is integrating our compilation and runtime techniques into OpenMP offload, which currently lacks task parallelism inside \texttt{target} regions.


\bibliographystyle{ACM-Reference-Format}
\bibliography{reference}

\end{document}